\documentclass[aip, jcp, twocolumn, amsmath, reprint]{revtex4-1}

\usepackage{graphicx}
\usepackage{color}
\usepackage{latexsym}
\usepackage{bm} 

\def\d {{\rm d}}
\def\tensor#1{{\mathbf{ {#1}}}}
\def\tensor#1{{\overline{\overline {#1}}}}
\def\div {{\vec\nabla\cdot}}

\def\tr{{\rm tr}}

\begin{document}

\title{Ultrafast spherulitic crystal growth as a stress-induced phenomenon\\ specific of fragile glass-formers}
\author{Christiane Caroli}
\affiliation{INSP, Universit\'e Pierre et Marie Curie-Paris 6, 
CNRS, UMR 7588, 4 place Jussieu, 75252 Paris Cedex 05, France}
\author{Ana\"el Lema\^{\i}tre}
\affiliation{Universit\'e Paris Est -- Laboratoire Navier, ENPC-ParisTech, LCPC, CNRS UMR 8205
2 all\'ee Kepler, 77420 Champs-sur-Marne, France}
\date{\today}

\begin{abstract}
We propose a model for the abrupt emergence, below temperatures close to the glass transition, of the ultra-fast (GC) steady mode of spherulitic crystal growth in deeply undercooled liquids. We interpret this phenomenon as controlled by the interplay between the generation of stresses by crystallization and their partial release by flow in the surrounding amorphous visco-elastic matrix. Our model is consistent with both the observed ratios ($\sim10^4$) of fast-to-slow velocities and the fact that fast growth emerges close to the glass transition. It leads us to conclude that the existence of a fast growth regime requires both (i) a high fragility of the glassformer; (ii) the fine sub-structure specific of spherulites. It finally predicts that the transition is hysteretic, thus allowing for an independent experimental test.
\end{abstract}

\maketitle

Upon approaching $T_g$, the rate of crystal growth from deeply supercooled glassformers decreases to very small values -- in the $10^{-12}$ m/s range -- as appears consistent with the dramatic slowing down of molecular motions. Growth is hence expected to be nearly arrested in the glass phase. This expectation, however, is challenged by a surprising phenomenon identified by Oguni and coworkers in several fragile materials~\cite{HikimaAdachiHanayaOguni1995,HataseHanayaOguni2004} and further investigated, more recently, by the groups of Yu and Ediger at Madison~\cite{ShtukenbergFreundenthalGunnYuKahr2011}. Namely below some temperature $T_t$ slightly larger than $T_g$, an ultrafast crystal growth mode (often coined GC, for Glass-Crystal) suddendly emerges. Near $T_t$ the ratio of the fast over normal (slow) front velocities, $u_f$ and $u_s$, is huge, typically of order $10^4$.

Up to now this phenomenon has been observed in 11 one-component molecular glass-formers~\cite{ShtukenbergFreundenthalGunnYuKahr2011}, which all have high fragility indices, that is, exhibit the so-called decoupling phenomenon~\cite{FujaraGeilSillescuFleischer1992,CiceroneEdiger1996}. Namely, below $\sim 1.2\,T_g$ the viscosity $\eta$ and translational diffusion coefficient $D$ of the SC liquid no longer obey the Stokes-Einstein relation but instead, $D\sim \eta^{-\xi}$, with $0<\xi<1$. Ediger et al~\cite{EdigerHarrowellYu2008} have shown that, as $T$ varies, the growth velocity of the slow (normal) mode scales, not as $\eta^{-1}$, but as $D$, a result consistent with the idea that crystallization kinetics is controlled by local processes~\cite{StillingerDebenedetti2005}. 

Moreover, materials exhibiting fast growth share the following common features: molecules are strongly anisotropic (nearly planar); single-crystals grown at small undercoolings are highly facetted -- which indicates slow interfacial kinetics; at large undercoolings, crystal growth is spherulitic~\cite{KeithPadden1963,BisaultRyschenkowFaivre1991} with, in the usual observation domain (spherulite radii $\sim$ 10's to 100's of microns), a quasi-constant front velocity.

To this day, the paradoxical emergence of this fast growth mode near $T_g$ largely remains a mystery. A noticeable proposition formulated by Tanaka~\cite{Tanaka2003} is that below $T_g$, volume contraction upon crystallization creates in the surrounding glassy matrix a negative pressure, i.e. free-volume; this would increase particle mobility close to the interface, hence accelerate crystallization. Konishi and Tanaka~\cite{KonishiTanaka2007} have shown that the values of the velocity ``jump'' at $T_t$ between fast and slow growth for two polymorphs of salol indeed correlate with their density changes upon crystallization. The Madison group, however, have later objected that such is not the case for the polymorphs of another material (ROY)~\cite{SunXiChenEdigerYu2008}. They moreover point out that a net accumulation of free-volume at the growing interface would result in accelerated growth, contrary to observations. Whether GC growth is a stress-induced phenomenon thus remains an open issue.

We will here uphold that such is indeed the case, provided that Tanaka's ideas are formulated in a slightly more general way.
Indeed, it seems physically undisputable that: (i) volume contraction upon crystallization creates stresses in the surrounding matrix and at the interface; (ii) these stresses modify interfacial kinetics. Now, we will argue that: (i) contraction does not result in net free-volume creation at the interface; (ii) however, in the absence of an average negative pressure, due to structural disorder, deviatoric interfacial stresses and pressure fluctuations are sufficient to accelerate growth. 

Besides, we note that GC growth appears near $T_g$, in a region where, as the relaxation time $\tau_\alpha$ increases, the mechanical response of the amorphous matrix rapidly crosses over from predominantly viscous (above $T_g$) to predominantly elastic. Its mechanical response is hence visco-elastic. The matrix rheology determines to which extent the elastic Eshelby stresses generated by spherulitic growth are relieved by the dissipative flow. 
The interfacial stress is thus controlled by the competition between the generation of stresses by growth and their relaxation by flow in the matrix.

In the following, we present a schematic model, which describes this interplay in a simple way, using the linear Maxwell rheology to model the matrix mechanical response. The model is developed in two stages. In its first ``idealized'' version, we represent the spherulite as an homogeneous, elastically isotropic, sphere. We then predict at low enough temperatures the coexistence of two growth modes:\\
-- a slow mode, which continously extrapolates from high temperatures, and in which interfacial stresses are almost fully relaxed by viscous flow in the matrix;\\
-- a fast one, in which, on the contrary, viscous stress relaxation is inefficient, so that stresses are essentially those produced by an Eshelby inclusion in an elastic matrix. \\
This model satisfactorily accounts for the order of magnitude (typically $\sim10^4$) of the ratio between the corresponding velocities, $u_f$ and $u_s$, yet also predicts that a ``fast spherulite'' would commute to slow growth past a $T$-dependent maximum radius which turns out to take unphysically small values (in the 100 nm range) around both $T_g$ and $T_t$, in contradiction with observations.

This leads us to question the above simplistic description of a spherulite. Indeed, as documented in detail by Faivre and coworkers~\cite{BisaultRyschenkowFaivre1991},  spherulites are compact objects which appear spherical on the optical scale, but are constituted, at smaller scales, of a space-filling accumulation of ``sheaves'' (of micrometric diameters), themselves constituted of anisotropic single crystals (typical size $\sim100$~nm), growing intermittently via small- and large-angle branching. Therefore, although spherulites of more than micrometric sizes can be viewed on a coarse-grained scale as spherical isotropic elastic inclusions in an amorphous matrix, crystallization actually proceeds via the intermittent growth of fiber-like units (the sheaves). This leads us to propose a second version of the model, which takes into account the fact that the growing units have a constant head radius comparable with a fiber diameter. This ``rugged spherulite'' model now predicts that fast growth emerges at low temperature via a hysteretic, saddle-node bifurcation, the existence of which is a direct consequence of the decoupling phenomenon. It accounts not only for the ratio of growth velocities, but also for the observability of fast growth up to the vicinity of $T_g$. Our interpretation of the GC-vs-normal growth phenomenon in terms of elastic vs viscous stress control moreover explains an important observation made by the Madison group, namely that the fast growth velocity is comparable with that of solid-solid transformations between polymorphs.

In Section~\ref{sec:1} we derive the equations which relate the interfacial stress and the front velocity for an elastic sphere growing from a visco-elastic matrix. In Section~\ref{sec:2} we show that such an idealized spherulite may present fast growth, but this only transiently, and up to maximal radii which are much too small to account for experimental observations. In Section~\ref{sec:3} we describe our modified, ``rugged spherulite'', model and show that it accounts for the emergence of steady fast growth at low temperatures via the crossing of a sub-critical, hysteretic, bifurcation.
In the discussion (Section~\ref{sec:4}), after summarizing our argument, we conclude that two physical ingredients are indispensable for the emergence GC growth, namely a high fragility of the supercooled liquid and a fine sub-structure of the growing crystal phase.

\section{Growth of a solid sphere from a visco-elastic medium}
\label{sec:1}

In the temperature range where GC growth occurs, close to the glass transition, crystal front velocities are at most of the order of $10^{-7}$m/s. The temperature rise $\delta T$ at the front due to latent heat production is of order $V\mathcal{L}R/K_m$~\cite{CaroliCaroliRouletFaivre1989}, with $V$ the front velocity, $R$ the spherulite radius, $\mathcal{L}$ the latent heat per unit volume, and $K_m$ the thermal conductivity of the amorphous matrix. With $\mathcal{L}\sim10^8$J/m$^3$, $K_m\sim 10^{-1}$J/(m.s.K), and for typical $R$ values $\lesssim200\mu$m~\cite{HataseHanayaOguni2004,SunXiChenEdigerYu2008}, we get $\delta T\lesssim 10^{-2}$K. We can therefore consider growth to occur in a completely isothermal matrix. 

In this section, we assume that the crystal is an isotropic elastic solid with a perfect spherical shape, growing from a mechanically isotropic, linear visco-elastic, amorphous matrix. Under the ambient pressure $p_0$, the matrix and crystal have equilibrium densities $\rho_0$ and $\rho_{c\,0}=\rho_0+\delta\rho_0$, with $\delta\rho_0>0$. That is we assume that the strain associated with crystallization is a pure volume contraction.

\subsection{Interfacial stress dynamics}

From a mechanical standpoint, we are dealing with an Eshelby-like problem~\cite{Eshelby1957}, as is clear if we consider, for a moment, the case when the matrix is a perfectly elastic solid. The stress field $\tensor\tau$ can be computed as follows: at any given instant $t$, a spherical volume $V_0$ of the matrix has transformed into a (crystalline) state characterized by a strain $\tensor\epsilon_0=-\frac{1}{3}\frac{\delta\rho_0}{\rho_0}\,\tensor 1$. Mechanical equilibrium then demands that $\nabla\cdot\tensor\tau=0$ inside each phase ($i=c,m$) with:
\begin{equation}\label{eq:sigma}
\tensor\tau_i=2\mu_i\,\tensor\epsilon_i+\lambda_i\,\tr\tensor\epsilon_i\,\,\tensor 1
\end{equation}
and 
\begin{equation}\label{eq:epsilon}
\tensor\epsilon_i=\frac{1}{2}\left(\vec\nabla \vec u_i+\vec\nabla\vec u_i^T\right)
\end{equation}
where $\tensor\epsilon_i$ is the elastic strain relative to the reference state of each phase (equilibrium under pressure $p$), and $\vec u_i$ the corresponding displacement. These equations must be supplemented with two interfacial conditions: (i) mechanical equilibrium, i.e. continuity of normal stresses; (ii) mass conservation on the moving front.

Since we are in spherical geometry under pure contraction, it is clear that:
\begin{equation}
\vec u_i= u_i(r) \vec e_r
\end{equation}
Mechanical equilibrium then entails that:
\begin{eqnarray}
\label{eq:um}
\text{in the matrix}\,\, (r>R): &\qquad u_m(r)=A_m/r^2\\
\label{eq:uc}
\text{in the crystal}\,\, (r<R): &\qquad u_c(r)= A_c r
\end{eqnarray}
It immediately follows that $\vec\nabla\cdot\vec u_m=0$ everywhere in the matrix, in particular at the interface.
This leads to a classical~\cite{Slaughter2001} but counterintuitive result: volume contraction upon crystallization does not give rise to any excess pressure in the surrounding matrix; it creates a purely deviatoric stress field.

The above discussion immediately carries over to the case of our concern of a linearly visco-elastic matrix, as the sole difference is that the Lam\'e coefficients $\lambda_m$ and $\mu_m$ now become frequency dependent. The space dependence of the displacement fields is still given by~(\ref{eq:um}) and~(\ref{eq:uc}) (with time-dependent coefficients $A_m$ and $A_c$). Hence, even in the presence of a dissipative mechanical response, the stresses remain purely deviatoric in the amorphous matrix. It follows that the viscous flow, if any, is also divergence free. Finally, no free-volume is created.

In the crystal $(r<R)$, the stress field (a pure pressure) is given by:
\begin{equation}
\tensor\tau_c(r,t)= (3K_c A_c-p_0)\,\tensor 1
\end{equation}
with $K_c=\lambda_c+\frac{2}{3}\mu_c$ the crystal bulk modulus. The crystal is under strain $A_c\,\tensor 1$, i.e. grows at density $\rho_{c\,0}+\delta\rho_c$ such that
\begin{equation}\label{eq:rhoc}
A_c=-\frac{1}{3}\frac{\delta\rho_{c}}{\rho_{c\,0}}
\end{equation}

In the matrix $(r>R)$, the strain field (purely deviatoric) is:
\begin{equation}
\tensor\epsilon_m(\vec r,t)=\frac{A_m}{r^3}(-2\vec e_r\vec e_r+\vec e_\theta\vec e_\theta+\vec e_\phi \vec e_\phi)
\end{equation}
Since we assume a linear Maxwell rheology the stress in the matrix is necessarily of the form:
\begin{equation}
\tensor\tau_m(\vec r,t)=\tau(r,t)(-2\vec e_r\vec e_r+\vec e_\theta\vec e_\theta+\vec e_\phi \vec e_\phi)-p_0\,\tensor 1
\end{equation}
where $\tau$ verifies:
\begin{equation}
O\star\tau(r,t)\equiv\left(\frac{1}{2\mu_m}\,\frac{\partial}{\partial t}+\frac{1}{2\eta}\right)\,\tau=\frac{1}{r^3}\frac{\d A_m}{\d t}
\end{equation}
with $\eta$ the viscosity of the amorphous phase. This solves at any point $r$ as $\tau(r,t)=\frac{1}{r^3}\,O^{-1}\star\dot A_m$ wherefrom the interfacial radial (tensile) stress $\Sigma(t)\equiv-2\tau(R(t),t)$ verifies $O\star(\Sigma\,R^3)=-2\dot A_m$, i.e.:
\begin{equation}\label{eq:sigma}
\frac{1}{2\mu_m}\,\frac{\d }{\d t}\left(\Sigma R^3\right)+\frac{\Sigma R^3}{2\eta}=-2\frac{\d A_m}{\d t}
\end{equation}

Mechanical equilibrium in the bulks of both phases must now be complemented by two interfacial conditions, namely, for our growth problem:\\
\emph{(i) continuity of normal stresses:}
\begin{equation}\label{eq:continuity}
3K_c A_c=\Sigma
\end{equation}
Note that we have omitted here the contribution of the interfacial tension $\gamma$. Indeed, it introduces a correction to the internal strain of the crystal of order $\frac{2\gamma}{K_cR}$. Taking $\gamma\sim10^{-2}$N/m, $K_c\gtrsim 10^{10}$Pa, and $R\gtrsim1\mu$m, this is at most of order $10^{-6}$, hence completely negligible.\\
\emph{(ii) mass conservation} across the interface moving at velocity $\dot R$:
\begin{equation}
\rho_0\,(\dot u_m(R)-\dot R)=\rho_c\,(\dot u_c(R)-\dot R)
\end{equation}
with, for any $r$, $\dot u_m(r)=\frac{{\dot A}_m}{r^2}$ and $\dot u_c(r)=\dot A_c r$, so that the above equation also writes:
\begin{equation}
\frac{\dot A_m}{R^3}=\left(\frac{\rho_0-\rho_{c\,0}-\delta\rho_c}{\rho_0}\right)\frac{\dot R}{R}+\frac{\rho_{c\,0}+\delta\rho_c}{\rho_0}\dot A_c
\end{equation}

\begin{figure*}[t]
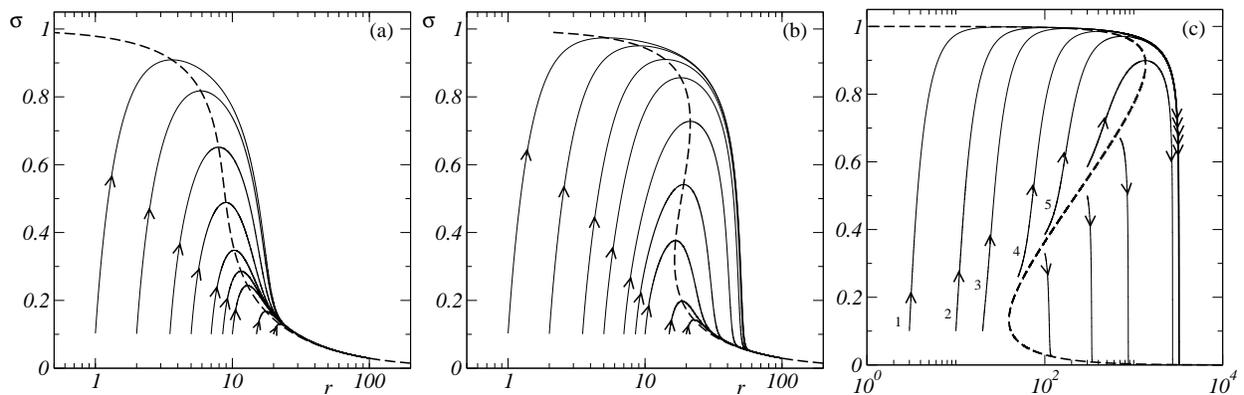

\includegraphics*[width=0.3\textwidth]{b_3.5.eps}
\includegraphics*[width=0.3\textwidth]{b_5.eps}
\includegraphics*[width=0.3\textwidth]{b_10.eps}
\caption{\label{fig:portrait}
Phase portrait of the dynamical system~(\ref{eq:reduced:all}) in the $(r,\sigma)$ plane, for $b=3.5$, 5, and 10 (from left to right).
}
\end{figure*}

The relative density change upon crystallization $\epsilon=\delta\rho_0/\rho_0$ is at most a few percent. Since we know from equation~(\ref{eq:rhoc}) that $A_c$ is $\mathcal{O}(\epsilon)$ we neglect the quadratic term $\frac{\delta\rho_c}{\rho_0}\dot A_c(R)$ in the above equation, whence:
\begin{equation}
\frac{\dot A_m}{R^3}=\left(-\epsilon+3A_c\right)\frac{\dot R}{R}+\dot A_c
\end{equation}
Together with equations~(\ref{eq:sigma}) and~(\ref{eq:continuity}) this now provides an equation relating the interfacial stress to both the growth velocity and the spherulite radius:
\begin{equation}\label{eq:stress}
{\dot\Sigma}+\left(\frac{3\dot R}{R}+\frac{\mu}{\eta}\right)\Sigma=4\mu\,\epsilon \,\frac{\dot R}{R}
\end{equation}
where we have defined an equivalent elastic modulus $\mu$ via:
\begin{equation}
\frac{1}{2\,\mu}=\frac{1}{2\,\mu_m}+\frac{2}{3K_c}
\end{equation}

\subsection{Interfacial kinetics}

Equation~(\ref{eq:stress}), which expresses both mechanical equilibrium and mass conservation conditions, must now be supplemented by the specification of interfacial kinetics. For a simple liquid, the crystallization process can be described as a series of activated jumps over barriers, of typical height $E$, separating the amorphous and crystalline molecular configurations. At very large undercooling, the probability of back jumps from the crystal into the matrix is negligible so that one may write the front velocity (in the absence of any interfacial stress) as $\sim\omega_0\,a\,e^{-\frac{E}{k_BT}}$, with $\omega_0$ a typical molecular frequency and $a$ an atomic length. In the fragile glassformers of interest here, molecular transport involves a wide, $T$-dependent, distribution $\rho(E;T)$ of barrier heights, so that one should schematically write:
\begin{equation}\label{eq:v0}
\dot R=V_0(T)\sim\omega_0\,a\,\int\,\d E\rho(E;T)\,e^{-\frac{E}{k_BT}}
\end{equation}
$V_0(T)$ is the growth velocity that should be measured in the absence of stress effects.

As discussed above, although the growth of an idealized spherulite produces no free-volume, it does generate finite shear stresses in the interfacial region. Like glasses, deeply undercooled liquids close above $T_g$ exhibit highly multistable free-energy landscapes. As elaborated long ago by Eyring, stresses bias such landscapes: a series of works~\cite{MalandroLacks1999,MaloneyLemaitre2006,RodneySchuh2009a,ChattorajCaroliLemaitre2011} has indeed now clearly shown that this effect controls the plastic deformation and rheology of amorphous solids. When an amorphous system is subjected to a given external shear stress field, the barrier limiting a particular local reconfiguration may increase or decrease depending on the relative orientation of molecular displacements with respect to the principal axes of the stress.

In the presence of interfacial shear stresses $\Sigma$, a barrier of height $E$ is thus shifted into $E+\delta E(\Sigma)$ and following Eyring, we assume that $\delta E=\alpha\Sigma$. Given the high structural disorder of the amorphous phase, we can reasonably consider that the $\alpha$'s have random signs. For the sake of simplicity, we furthermore assume that they are sharply peaked around $\pm \alpha_0$, with $\alpha_0$ a typical ``activation volume''. The growth front velocity now becomes:
\begin{equation}\label{eq:rdot}
\dot R=V_0(T)\cosh\left(\frac{\alpha_0\Sigma}{k_BT}\right)
\end{equation}
where $V_0(T)$ is given by equation~(\ref{eq:v0}).
Our growth problem is now fully specified by equations~(\ref{eq:stress}) and~(\ref{eq:rdot}).

\section{Growth of an idealized spherulite}
\label{sec:2}

In this section, we analyze the solutions of the above growth model. That is, we solve equations~(\ref{eq:stress}) and~(\ref{eq:rdot}) starting from initial conditions corresponding to an already formed idealized spherulite well beyond the nucleation regime. Since we are primarily interested in the transition between normal and GC growth, we restrict our attention to temperatures close to $T_g$.

We first rewrite the growth equations in a dimensionless form, by introducing the following units for, respectively stress, length, and time:
\begin{equation}\label{eq:scales}
\Sigma_0 = \frac{4\epsilon\mu}{3}\quad,\qquad \lambda = \frac{\eta}{\mu}\,V_0\quad\text{and}\qquad \tau=\frac{\eta}{\mu}
\end{equation}
Note that, in contrast with the stress scale $\Sigma_0$, both $\lambda$ and $\tau$ depend on temperature via the rapidly varying quantities $\eta$ and $V_0$.
Denoting
\begin{equation}
\sigma = \frac{\Sigma}{\Sigma_0}\qquad \text{and}\qquad r = \frac{R}{\lambda}
\end{equation}
equations~(\ref{eq:stress}) and~(\ref{eq:rdot}) now become:
\begin{subequations}\label{eq:reduced:all}
\begin{eqnarray}
\dot\sigma+\sigma &=& \dfrac{3\,\dot r}{r}\,(1-\sigma)\label{eq:reduced:a}\\
\dot r &=& \cosh(b\,\sigma)\label{eq:reduced:b}
\end{eqnarray}
\end{subequations}
The reduced dynamics is thus controlled by a single parameter
\begin{equation}\label{eq:b}
b=\frac{\alpha_0\Sigma_0}{k_B T}
\end{equation}
which measures the sensitivity of interfacial kinetics to tensile stress.

The behaviour of the solutions of the dynamical system~(\ref{eq:reduced:all}) is best summed up by plotting its phase portrait in the $(r,\sigma)$ plane, as shown on Fig.~\ref{fig:portrait} for three values of $b$. The $\dot\sigma=0$ condition, which is verified when:
\begin{equation}
r=r^\star(\sigma)=3\dfrac{1-\sigma}{\sigma}\cosh(b\sigma)
\end{equation}
defines a curve, denoted $C^\star$, in the $(r,\sigma)$ plane (dashed lines on Fig~\ref{fig:portrait}).
On the left ($r< r^\star(\sigma)$) of $C^\star$, the interfacial stress increases as growth proceeds. On its right ($r> r^\star(\sigma)$), $\sigma$ decreases until it approaches asymptotically the low-$\sigma$, high-$r$ part of $C^\star$. 

Starting from a small initial $r$ value, the stress $\sigma$ first increases very sharply. For small $b$'s, it quickly drops back to very small values after reaching its maximum (at the crossing point with $C^\star$).

As $b$ increases beyond $b_c\simeq4$, $C^\star$ develops an increasingly marked reentrance with a flat upper branch $\sigma\sim1$ extending up to a value $r_{\rm max}$ that grows exponentially with $b$. The trajectories stay close to this high-$\sigma$ branch up to radii $> r_{\rm max}$, then drop sharply to reach the flat $\sigma\ll1$ branch. Note that for $b\gg b_c$ (see Fig~\ref{fig:portrait}c for $b=10$), as most trajectories originating from the left of $C^\star$ (see curves 1 to 5 on the figure) come closely together onto the common high-$\sigma$ plateau, they also drop at about the same value $r_{\rm switch}$. 

Since [see Eq.~(\ref{eq:reduced:b})] $\sigma$ measures the logarithm of the growth velocity, the high-$\sigma$ plateau corresponds to a transient fast growth regime extending over a range of radii which, although finite, increases exponentially with $b$. It now appears natural to check whether we can interpret the upper and lower branches of $C^\star$ as corresponding to the GC and normal spherulitic growth regimes respectively. 
For this purpose, we must be able to account first of all for the two main experimental features: (i) the order of magnitude of the ratio of the fast to normal growth velocities in the vicinity of $T_g$; (ii) the existence of an upper temperature $T_t\gtrsim T_g$ above which compact GC growth is not observed.

Comparisons with experiments are here performed on the especially well documented case of OTP ($o$-terphenyl). For this material, at $T_g=243$K~\cite{SunXiEdigerYu2008}, where the viscosity $\eta=10^{11}$Pa.s, the contraction parameter $\epsilon\approx4.8\%$~\cite{NaokiKoeda1989}; using the longitudinal wave speed $c_l=2550$m/s~\cite{BuchenauWischnewski2004} and the ratio $c_l/c_t\sim2$~\cite{NovikovSokolov2004} we estimate a shear modulus value $\mu\sim1.7$GPa; the molar mass is $M=230$g/mol; and the specific volume $v=0.89\,$cm$^3$/g~\cite{PlazekBeroChay1994}. 

\emph{(i) Velocity ratio.} In our idealized spherulite model, the normal growth velocity ($\sigma\ll1$) is $u_s\sim V_0$. Fast growth ($\sigma\sim1$) occurs under an interfacial tension $\sim\Sigma_0=\frac{4\epsilon\mu}{3}$, corresponding to a shear stress $\sim\Sigma_0/2=\frac{2\epsilon\mu}{3}\sim 3.10^{-2}\mu$ -- that is, a sizeable fraction of the yield stress. The growth velocity $u_f\sim V_0\cosh b$. From the measured value $u_f/u_s\approx10^{4}$, we deduce that $b\approx10$.

A first assessment of whether fast growth may result from the stress-induced acceleration of interfacial kinetics consists in checking that these parameters correspond to a reasonable value of the activation volume $\alpha_0$. With the help of Eq.~(\ref{eq:b}), we find $\alpha_0\approx0.3\,$nm$^3$, precisely of the order of the molecular volume $v_m\approx 0.34\,$nm$^3$, which indeed brings valuable support to this idea.

\emph{(ii) Switching from GC to normal growth.} Our idealized spherulite model predicts that GC growth (when initial conditions allow for its occurrence) cannot be a permanent regime, but must switch to normal growth when the reduced radius reaches an upper limit $r_{\rm switch}$.
Can we reconcile this prediction with the experimental observation of a temperature $T_t$ above which compact GC growth ceases to be observed? Let us note that the published data are concerned with spherulite radii of a few hundreds of micrometers at most, hence do not necessarily contradict the possibility that GC growth be transient.

The upper physical radius at which GC growth is predicted to be observable is
\begin{equation}
R_{\rm switch}=\lambda\,r_{\rm switch}(b)
\end{equation}
where [see Eq.~(\ref{eq:scales})] $\lambda=\eta\,V_0/\mu$.
As shown by the Madison group~\cite{SunXiEdigerRichertYu2009} materials where GC growth is observed have high fragilities and, as such, exhibit a decoupling between translational diffusion and viscosity~\cite{CiceroneEdiger1996,AngellNgaiMcKennaMcMillanMartin2000,DebenedettiStillinger2001}. Namely, below about $1.2\, T_g$ the translational diffusion coefficient $D$ no longer scales as $\eta^{-1}$, but as $\eta^{-\xi}$ with $0<\xi<1$. As emphasized in~\cite{SunXiEdigerRichertYu2009}, these materials have comparable values of $\xi$ of order 0.7. Moreover, Ediger \emph{et al.}~\cite{EdigerHarrowellYu2008} have shown that the normal growth velocity $V_0$ scales, not as $\eta^{-1}$, but as $D$ -- a result consistent with the idea that crystallization kinetics is controlled by local processes. In the narrow vicinity of $T_g$ we are interested in, the equivalent shear modulus $\mu$ is quasi-constant as compared with the fast-varying quantity $\eta\,V_0$ and our lengthscale $\lambda$ scales with $T$ as:
\begin{equation}\label{eq:lambda}
\lambda\sim\eta\,D\sim\eta^{1-\xi}
\end{equation}

As temperature increases, $b$ [see Eq.~(\ref{eq:b})], hence $r_{\rm switch}(b)$, decreases and so does $\lambda$. At a fixed temperature GC growth would thus be observable up to a maximum radius $R_{\rm switch}$, which decreases with $T$. We must now examine whether our idealized model is compatible with the observation of spherulites of radii $\sim$ a few $100\mu$m at $T_g$. For $b=10$, we find $r_{\rm switch}\simeq 3.10^3$; from~\cite{SunXiEdigerYu2008} we estimate for OTP $V_0\sim10^{-12}$m/s; with the above values of $\eta$ and $\mu$, this yields: $R_{\rm switch}\sim0.2\mu$m. Clearly, the idealized spherulite model, as such, fails to account for experimental observations.

Does this failure invalidate our primary assumption that GC is a stress-induced phenomenon? Or could it be attributed to shortcomings in our extremely idealized representation of a spherulite?

It is useful at this stage to return to Eq.~(\ref{eq:stress}) and note that when stress relaxation in the amorphous matrix can be neglected, that is for $\mu/\eta=0$, the interfacial stress asymptotes to the stationary value $\Sigma_0=4\epsilon\mu/3$~\footnote{The fact that interfacial stress does not increase indefinitely as growth proceeds is consistent with the absence of free-volume creation.}, which is precisely the stress level at which transient GC growth occurs in the above description. This means that, in the stress-controlled model, GC growth is too fast for stress relaxation in the supercooled matrix to be efficient and occurs at a velocity close to that of a crystal growing from a solid matrix. This qualitative result must be put in regard with an important remark made by the Madison group~\cite{SunXiChenEdigerYu2008}: namely, they point out that the order of magnitude of GC growth velocities compares with that of solid-solid transformations between polymorphs of the same material.

In our opinion, this, together with the internal consistency of our prediction for the velocity ratio, makes a strong case for retaining the stress-controlled assumption and reconsidering the details of our model.

\section{Growth of a ``rugged spherulite''}
\label{sec:3}

The representation we have used so far of a spherulite as a perfectly spherical homogeneous solid was drastically over-simplified. Indeed, these objects result from the growth of anisotropic micro-crystals which, as they multiply via intermittent small- and large-angle branching, form a space-filling quasi-spherical polycrystalline structure. So, as seen on Fig.6b of reference~\cite{BisaultRyschenkowFaivre1991}, the crystal-matrix interface is rugged on a scale in the $\sim100$~nm range, fixed by the micro-crystal growth habit, and thus independent of the average spherulite radius $R$. Clearly, ahead of such an interface, the stress field must be modulated on the same scale. 

We must therefore reformulate our growth model to take into account the presence of this interfacial ruggedness. Our model boils down to two equations: (i) the kinetic equation~(\ref{eq:rdot}) which relates the growth velocity to the interfacial stress; (ii) the mechanical equation~(\ref{eq:stress}) which describes the evolution of interfacial stress under the combined effects of growth and viscous flow in the matrix.
The kinetic equation, which is local, hence independent of the inclusion shape, clearly remains unchanged.
The mechanical equation, however, which was derived in Section~\ref{sec:1} for the special case of a perfect sphere, must now be reconsidered.

For this purpose, we first reinterpret equation~(\ref{eq:stress}) in the simple, fully elastic, limit (${\mu}/{\eta}=0$). 
We denote $\mathcal{C}(t)$ the domain occupied by the crystal at time $t$, and $\mathcal{I}(t)$ the corresponding interfaces. Let us assume that at time $t_1$, the stress field in the system satisfies the conditions of mechanical equilibrium ($\nabla\cdot\tensor\tau=0$ everywhere; continuity of normal stress on $\mathcal{I}(t_1)$). The change of the stress field due to spherulitic growth over the interval $[t_1,t_2]$ is due to the amorphous-to-crystal (A $\to$ C) transformation of the domain extending between $\mathcal{C}(t_1)$ and $\mathcal{C}(t_2)$; thanks to linear superposition, this is  equivalent to the following sequence:\\
1) C $\to$ A transformation of $\mathcal{C}(t_1)$\\
2) A $\to$ C transformation of $\mathcal{C}(t_2)$\\
In other words, inside the matrix, at any point:
\begin{equation}\label{eq:eshesh}
\tensor\tau(\vec r,t_2)-\tensor\tau(\vec r,t_1)=
\tensor\tau_{\rm Esh.}(\vec r,t_2)-\tensor\tau_{\rm Esh.}(\vec r,t_1)
\end{equation}
where $\tau_{\rm Esh.}(\vec r,t)$ is the Eshelby field~\cite{Eshelby1957,Slaughter2001} due to the A $\to$ C transformation of an inclusion $\mathcal{C}(t)$.

Consider now a point $\vec r(t)$, which lies on the interface at all times, and is constrained to move along the local normal vector 
(${\d \vec r}/{\d t}=(\vec V\cdot\vec n) \vec n$). Using~(\ref{eq:eshesh}), the interfacial stress at this running point, $\tensor\tau(\vec r(t), t)$, is found to verify:
\begin{equation}\label{eq:interfacial}
\frac{\d\tensor\tau(\vec r(t), t)}{\d t}=(\vec V\cdot\vec n)\,\vec n\cdot\vec\nabla\left(\tensor\tau-\tensor\tau_{\rm Esh.}\right)+
\frac{\d\tensor\tau_{\rm Esh.}(\vec r(t), t)}{\d t}
\end{equation}

Coming back to the case of a growing sphere, for which $\tensor\tau_{\rm Esh.}(\vec r(t), t)={\rm Cst}$, let us rewrite equation~(\ref{eq:stress}), which governs the dynamics of interfacial stress, as:
\begin{equation}\label{eq:stress:2}
\dot\Sigma+\frac{\mu}{\eta}\,\Sigma=\frac{3\dot R}{R}(\Sigma_0-\Sigma)
\end{equation}
The second term on the l.h.s of this equation results from our assumed Maxwell rheology in the matrix. 
Comparison with Eq.~(\ref{eq:interfacial}) then shows that the factor $\dot R/R$ in the r.h.s. must be understood as the ratio of the normal growth velocity to the characteric length scale of stress gradients ahead of the growing front.

Clearly, given the complexity of spherulitic growth~\cite{BisaultRyschenkowFaivre1991}, a detailed description of the rugged front geometry as well as of the map of local normal velocities remains for the moment out of reach. Let us observe, however, that, as mentioned above, a spherulite can be viewed as a compact sphere of radius $R$ constituted of space-filling substructures of typical lateral dimension $D\sim100$nm (we refer here to the scale denoted $\epsilon$ in Ref.~\cite{BisaultRyschenkowFaivre1991}).
We evaluate in the appendix the stress field near the surface of a spherical inclusion decorated with bulges of typical size $D<<R$. We find that: (i) the tensile stress at the interface remains nearly equal to its value $\Sigma_0$ for the ideal sphere; (ii) the dominant length scale controlling the stress gradient is the typical bulge size $D$.

On the basis of these remarks, we hypothesize that the typical interfacial stress is governed by an equation similar to~(\ref{eq:stress:2}), yet in which the factor $\dot R/R$ is replaced by $V_n/D=\dot R/D$, where $D$ is a constant characteristic of the spherulitic micro-structure:
\begin{equation}\label{eq:stress:rugged}
\dot\Sigma+\frac{\mu}{\eta}\,\Sigma=\frac{3\dot R}{D}(\Sigma_0-\Sigma)
\end{equation}
while the kinetic equation~(\ref{eq:rdot}) remains unchanged.
In the reduced units defined in~(\ref{eq:scales}), and after elimination of $\dot R$, the dynamical problem reduces to:
\begin{equation}\label{eq:stress:rugged}
\dot\sigma+\sigma=\frac{3\lambda}{D}(1-\sigma)\,\cosh(b\sigma)
\end{equation}
which is similar to Eq.(\ref{eq:reduced:a}), with the difference that the dynamical variable $r$ is now replaced by the constant 
\begin{equation}
d\equiv D/\lambda
\end{equation}

\begin{figure}[h]
\includegraphics*[width=0.45\textwidth]{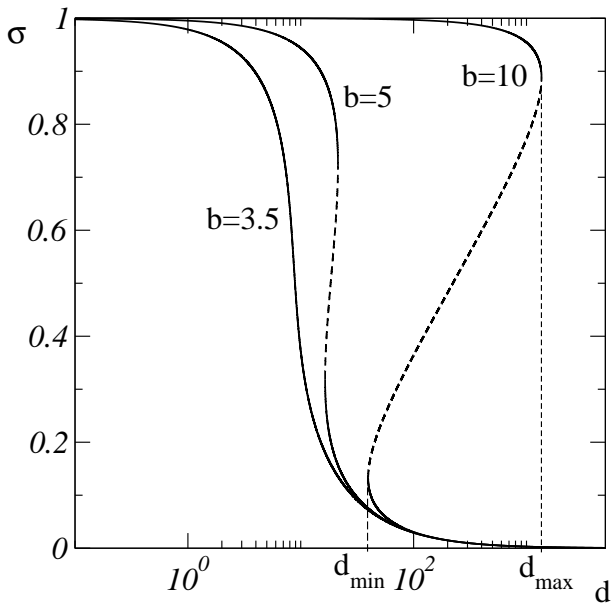}
\caption{\label{fig:2}
Loci of the fixed points of Eq.~(\ref{eq:stress:rugged}) in the $(d,\sigma)$ plane, for $b=3.5$, 5, and 10. Dashed lines: unstable branches.}
\end{figure}

In contrast with the ideal spherical growth model [Eq.~(\ref{eq:reduced:all})], the dynamical problem~(\ref{eq:stress:rugged}) presents fixed points, which are defined by the condition $\dot\sigma=0$, or equivalently by:
\begin{equation}
d=3\dfrac{1-\sigma}{\sigma}\cosh(b\sigma)
\end{equation}
The loci of these fixed points in the $(d,\sigma)$ plane, $\sigma^*(d)$, are the same as the dashed curves shown on Fig.~\ref{fig:portrait} and are reproduced on Fig.~\ref{fig:2}. The fixed points are stable when $\d\sigma^*/\d d<0$. When $b>b_c\simeq4$, as shown in Section~\ref{sec:2}, the curves $\sigma^*$ develop a reentrant branch, which is thus unstable. In this case, the nearly flat high- (resp. low-) $\sigma^*$ branches, corresponding to stable steady growth modes, present end points for $D/\lambda= d_{\rm max}$ (resp. $d_{\rm min}$). If $D/\lambda>d_{\rm max}$, the only stable growth mode is the $\sigma<<1$, slow one; analogously, for $D/\lambda<d_{\rm min}$, the only steady growth mode is the $\sigma\sim1$, fast one. Between these two values, depending on the initial conditions, the system settles in either of these two regimes.

As discussed at the end of Section~\ref{sec:2}, when temperature $T$ increases, $b$, hence $d_{\rm max}$, decrease, while due to the decoupling effect $\lambda$ [see Eq.~(\ref{eq:lambda})] decreases: $\lambda\,d_{\rm max}$ thus decreases with $T$. Moreover, as shown in Ref.~\cite{BisaultRyschenkowFaivre1991}, $D$ exhibits a weak increase in the 100nm range. Consequently, the fast growth mode exists only up to a maximum temperature $T_{\rm max}$ such that: 
\begin{equation}
\left.\frac{D}{\lambda}\right|_{T=T_{\rm max}}=d_{\rm max}(T_{\rm max})
\end{equation}
Similarly, steady slow growth exists only above the temperature $T_{\rm min}<T_{\rm max}$ such that: $D/\lambda =d_{\rm min}$.

Our rugged spherulite model thus predicts the existence of a finite temperature range where both slow or fast growth can be observed. Therefore, we expect the following hysteretic behaviour: after a quench at very low temperature, a spherulite displays steady fast growth; as temperature is slowly increased, the front velocity remains quasi-constant before reaching the vicinity of $T_{\rm max}$ where it slightly decreases; at $T=T_{\rm max}$, growth commutes to the slow (normal) mode~\footnote{The characteristic time scale for the transient dynamics is of course the matrix relaxation time $\tau_\alpha$}. Conversely, a slowly growing spherulite should commute to fast growth below $T=T_{\rm min}<T_{\rm max}$. The OTP data on Fig.~1 of Ref.~\cite{SunXiEdigerYu2008} indeed seem to present a finite (narrow) range of temperatures where both growth modes are observed.

The value of the fast-to-slow velocity ratio $u_f/u_s\sim\cosh(b)$ is the same as for the ideal spherulite. As seen in Section~\ref{sec:2}, the experimental value $u_f/u_s\sim10^4$ for OTP, leads to $b\approx10$ consistent with a reasonable estimate for the activation volume $\alpha_0$. We now examine whether our model is compatible with the experimental observation that the fast-to-slow transition occurs close to $T_g$. For $b=10$, we find $d_{\rm max}\simeq 1500$; with the above estimates at $T_g$ for OTP ($V_0\sim10^{-12}$m/s, $\eta/\mu\sim60$s) this leads to $D=V_0\,d_{\rm max}\,\eta/\mu\approx100$nm, a quite satisfactory estimate.

\section{Discussion}
\label{sec:4}
On the basis of the above analysis, we propose that GC growth is a stress-induced phenomenon.
Our argument can be summarized as follows. \\
(i) Since the crystal phase is denser than the amorphous one, the crystallization process generates stresses, in particular along the growth front. The growing spherulite can be viewed as an Eshelby inclusion, and as such, due to its globally spherical shape, does not create any net free-volume change in the matrix. However, a shear stress $\Sigma$ alone is able to accelerate interfacial kinetics, with respect to the normal diffusion-controlled regime, by a factor which we roughly estimate as $\sim\cosh(\Sigma\alpha_0/k_BT)$, with $\alpha_0$ an Eyring-like activation volume, which must be of order a molecular volume. With interfacial stresses of order of the Eshelby stress for a sphere $\Sigma_0=\frac{4}{3} \mu \delta\rho_0/\rho_0$, it immediately appears that moderate volume contractions of order a few $\%$ are sufficient to amplify growth velocity by several decades ($\sim10^4$ in the case of OTP). Moreover, this relation between the level of Eshelby interfacial shear stresses generated by the phase change and the amplification factor values is fully consistent with the observation by the Madison group~\cite{SunXiChenEdigerYu2008} that the velocity of GC growth compares with that of solid-solid transformations between polymorphs of the same material.\\
(ii) Since the matrix is visco-elastic, it is able to partially relax the stresses generated by crystal growth. From this viewpoint slow growth is possible only as long as stress relaxation is much faster than stress generation. As discussed in the previous section, ahead of the growth front, stresses decrease on a characteristic length scale $\ell$ which is fixed by the local geometry of the crystal. The characteristic time scale of stress variation in the matrix ahead of a front moving at velocity $V_0$ is thus $\tau\sim\ell/V_0$. Slow growth, which is the rule above $\sim1.2\,T_g$, can hence be sustained so long as this time scale is larger than $\tau_\alpha$, i.e.:
\begin{equation}
V_0\tau_\alpha\lesssim\ell
\end{equation}
Note that since $V_0$ and $\tau_\alpha$ scale respectively as $D$ and $\eta$, $V_0\tau_\alpha\sim\eta\,D$. For strong glassformers, the Stokes-Einstein scaling $\eta\,D\sim T$ holds; $V_0\tau_\alpha$ thus decreases with temperature: commutation from slow to fast growth cannot occur upon cooling. It is only in fragile glassformers, where below $T\sim1.2\,T_g$, $\eta\,D\sim\eta^{1-\xi}$ (with $0<\xi<1$) increases rapidly upon cooling, that the emergence of GC growth is possible.\\
(iii) The last question is then to decide on the order of magnitude of the length scale $\ell$. We have shown in Section~\ref{sec:2} that modelling a spherulite as a compact homogeneous sphere of radius $R$ is too naive: since it yields $\ell\sim R$, GC growth can only be a transient phenomenon; moreover, close to $T_g$, where GC growth is found to emerge, the maximum radius of a fast spherulite would then lie in the sub-micrometric range, contrary to observations. We have then argued that the relevant length scale is the width $D$ of the elementary substructures produced by the intermittent branching growth process, specific of spherulitic growth. We have then shown in Section~\ref{sec:3} that the order of magnitude $D\sim100$nm is consistent with the observed value of the emergence temperature $T_t\simeq T_g$. 

On this basis, we claim that \emph{two conditions are necessary} for the emergence of steady GC growth: (i) the existence of the \emph{decoupling phenomenon}~\cite{FujaraGeilSillescuFleischer1992,CiceroneEdiger1996} in the undercooled liquid (i.e. a large fragility index); (ii) a \emph{fine sub-structure} of the growing crystal (i.e. a rugged interface), which is the case of spherulites~\cite{BisaultRyschenkowFaivre1991}.

At this stage, our model is consistent with the main experimental facts. Moreover, one of its outcomes, namely the prediction that the slow to GC growth ``transition'' is hysteretic, should provide an independent test of its validity. Experimental investigation of this question, if feasible, would be highly desirable.

As a last remark, let us emphasize that our rugged spherulite model remains oversimplified, in particular as it assumes that the interfacial sub-structures share a common length scale $D$. In practice, in an actual spherulite, the size of sub-structures and growth velocities of the corresponding protrusions fluctuate. Consequently, the temperature at which GC commutes to normal growth may fluctuate along the growth front. This might explain the important observation~\cite{SunXiEdigerYu2008,SunXiChenEdigerYu2008} that at temperatures between $T_t$ and $\sim1.2\,T_g$, sparse fast growing fibers emerge from the slow growth fronts with which they coexist.

\appendix

\section{Appendix}

We recall here a few results about the classical Eshelby problem. Within an infinite elastically isotropic medium, we consider a domain $\mathcal{C}$ (the inclusion) that undergoes a density change as it transforms into another phase with the same elastic constants, under the condition of continuity of displacement at the interface. Since the transformation is a pure compression the total resulting strain field can be written as~\cite{Eshelby1957,Slaughter2001}:
\begin{equation}\label{eq:eps:general}
\tensor\epsilon=\frac{(1+\nu)\,\epsilon}{12\pi\,(1-\nu)}\,\vec\nabla\vec\nabla\phi
\end{equation}
with $\epsilon=\frac{\delta\rho_0}{\rho_0}$ the relative density change and with $\phi$ the elastic potential defined as
\begin{equation}\label{eq:phi:general}
\phi(\vec r)=\int_{\mathcal{C}}\d\vec r'\frac{1}{|\vec r-\vec r'|}
\end{equation}
which verifies
$\Delta\phi = 4\,\pi$ in the inclusion and $\Delta\phi=0$ outside.

This last equation entails that $\div\tensor\epsilon=0$ in the surrounding matrix. That is, whatever the shape of the inclusion, the transformation generates neither free-volume nor pressure field.

\begin{figure}[h]
\begin{center}
\includegraphics*[width=0.25\textwidth]{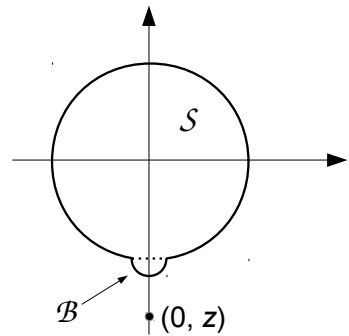}
\end{center}
\caption{\label{fig:bulge} A spherical inclusion with a small hemispherical bulge.}
\end{figure}

Moreover, given a partitioning of the inclusion, it immediately results from the form of equations~(\ref{eq:eps:general}) and~(\ref{eq:phi:general}) that the strain $\tensor\epsilon$ can be written as the sum of the contributions of different sub-domains. 
In the case illustrated on Fig.~\ref{fig:bulge} of an inclusion formed of a sphere $\mathcal{S}$ of radius $R$ decorated on its surface by a bulge $\mathcal{B}$, the stress in the matrix is thus written as $\tensor\sigma=2\mu\,(\tensor\epsilon_{\mathcal{S}}+\tensor\epsilon_{\mathcal{B}})$ with $\tensor\epsilon_{\mathcal{S}}$ and $\tensor\epsilon_{\mathcal{B}}$ the strains generated by the sphere and bulge respectively.

We note that for a inclusion with rotational symmetry about the $z$ axis, and defined in cylindrical coordinates by $\rho<\rho_{\mathcal{C}}(z)$ with $z\in[z_{\rm min},z_{\rm max}]$, the $zz$ component of the stress at a point $\vec r=(0,0,z)$ reads:
\begin{equation}\label{eq:integral}
\sigma_{zz}=\frac{\mu\,\epsilon}{3}\,\frac{1+\nu}{1-\nu}\,\int_{z_{\rm min}}^{z_{\rm max}}\d z'\left[\frac{{\rho'}^2}{\left({\rho'}^2+(z-z')^2\right)^{3/2}}\right]_0^{\rho_{\mathcal{C}}(z')}
\end{equation}

We now specialize to the case of a hemispherical bulge of radius $D\ll R$, in which case we can consider the inclusion as combining a full sphere with a hemisphere. For a sphere of radius $R$ centered at the origin one immediately recover from~(\ref{eq:integral}) the classical result:
$$
\sigma_{zz}^{\mathcal{S}}(z)=\sigma^{0}\left(\frac{R}{|z|}\right)^3\qquad {\rm for}\; |z|>R
$$
with $\sigma^{0}=\frac{4\mu\,\epsilon}{9}\,\frac{1+\nu}{1-\nu}$ the value of the tensile stress at the surface of the sphere.
For a hemisphere of radius $D$ centered at $z=-R$, we find:
$$
\sigma_{zz}^{\mathcal{B}}(z)=c\,\sigma^0\left(\frac{D}{|z+R|}\right)^3
\qquad {\rm for}\; z<-R-D
$$
with the prefactor $c=1-\frac{5}{4\sqrt{2}}\simeq0.12$.
At the interface, $z=-R-D$, the tensile stress is thus $((1+D/R)^{-3}+c)\,\sigma^0$.

The presence of a small bulge hence only slightly modifies the interfacial stress as compared with the value $\sigma^0$ for a perfect sphere.
However, close to the interface (up to distances of order $D$ at least), the stress gradient is dominated by the contribution of the bulge, that is, the length scale of the stress variations is the bulge radius $D$.

\acknowledgements

We are very grateful to Hajime Tanaka for attracting our attention to the puzzle of GC growth. 


\begin{thebibliography}{10}%
\makeatletter
\providecommand \@ifxundefined [1]{%
 \ifx #1\undefined \expandafter \@firstoftwo
 \else \expandafter \@secondoftwo
\fi
}%
\providecommand \@ifnum [1]{%
 \ifnum #1\expandafter \@firstoftwo
 \else \expandafter \@secondoftwo
\fi
}%
\providecommand \enquote [1]{``#1''}%
\providecommand \bibnamefont  [1]{#1}%
\providecommand \bibfnamefont [1]{#1}%
\providecommand \citenamefont [1]{#1}%
\providecommand\href[0]{\@sanitize\@href}%
\providecommand\@href[1]{\endgroup\@@startlink{#1}\endgroup\@@href}%
\providecommand\@@href[1]{#1\@@endlink}%
\providecommand \@sanitize [0]{\begingroup\catcode`\&12\catcode`\#12\relax}%
\@ifxundefined \pdfoutput {\@firstoftwo}{%
 \@ifnum{\z@=\pdfoutput}{\@firstoftwo}{\@secondoftwo}%
}{%
 \providecommand\@@startlink[1]{\leavevmode}%
 \providecommand\@@endlink[0]{}%
}{%
 \providecommand\@@startlink[1]{%
  \leavevmode
  \pdfstartlink
   attr{/Border[0 0 1 ]/H/I/C[0 1 1]}%
   user{/Subtype/Link/A<</Type/Action/S/URI/URI(#1)>>}%
  \relax
 }%
 \providecommand\@@endlink[0]{\pdfendlink}%
}%
\providecommand \url  [0]{\begingroup\@sanitize \@url }%
\providecommand \@url [1]{\endgroup\@href {#1}{\urlprefix}}%
\providecommand \urlprefix [0]{URL }%
\providecommand \Eprint[0]{\href }%
\@ifxundefined \urlstyle {%
  \providecommand \doi [1]{doi:\discretionary{}{}{}#1}%
}{%
  \providecommand \doi [0]{doi:\discretionary{}{}{}\begingroup
  \urlstyle{rm}\Url }%
}%
\providecommand \doibase [0]{http://dx.doi.org/}%
\providecommand \Doi[1]{\href{\doibase#1}}%
\providecommand \selectlanguage [0]{\@gobble}%
\providecommand \bibinfo [0]{\@secondoftwo}%
\providecommand \bibfield [0]{\@secondoftwo}%
\providecommand \translation [1]{[#1]}%
\providecommand \BibitemOpen[0]{}%
\providecommand \bibitemStop [0]{}%
\providecommand \bibitemNoStop [0]{.\EOS\space}%
\providecommand \EOS [0]{\spacefactor3000\relax}%
\providecommand \BibitemShut [1]{\csname bibitem#1\endcsname}%
\bibitem{HikimaAdachiHanayaOguni1995}%
  \BibitemOpen
  \bibfield{author}{%
  \bibinfo {author} {\bibfnamefont{T.}~\bibnamefont{Hikima}}, \bibinfo {author}
  {\bibfnamefont{Y.}~\bibnamefont{Adachi}}, \bibinfo {author}
  {\bibfnamefont{M.}~\bibnamefont{Hanaya}},\ and\ \bibinfo {author}
  {\bibfnamefont{M.}~\bibnamefont{Oguni}},\ }%
  \bibfield{journal}{%
  \Doi{10.1103/PhysRevB.52.3900}{\bibinfo {journal} {Physical Review B}}\ }%
  \textbf{\bibinfo {volume} {52}},\ \bibinfo {pages} {3900} (\bibinfo {month}
  {Aug.}\ \bibinfo {year} {1995})\BibitemShut{NoStop}%
\bibitem{HataseHanayaOguni2004}%
  \BibitemOpen
  \bibfield{author}{%
  \bibinfo {author} {\bibfnamefont{M.}~\bibnamefont{Hatase}}, \bibinfo {author}
  {\bibfnamefont{M.}~\bibnamefont{Hanaya}},\ and\ \bibinfo {author}
  {\bibfnamefont{M.}~\bibnamefont{Oguni}},\ }%
  \bibfield{journal}{%
  \Doi{10.1016/j.jnoncrysol.2003.10.010}{\bibinfo {journal} {Journal of
  Non-crystalline Solids}}\ }%
  \textbf{\bibinfo {volume} {333}},\ \bibinfo {pages} {129} (\bibinfo {month}
  {Feb.}\ \bibinfo {year} {2004})\BibitemShut{NoStop}%
\bibitem{ShtukenbergFreundenthalGunnYuKahr2011}%
  \BibitemOpen
  \bibfield{author}{%
  \bibinfo {author} {\bibfnamefont{A.}~\bibnamefont{Shtukenberg}}, \bibinfo
  {author} {\bibfnamefont{J.}~\bibnamefont{Freundenthal}}, \bibinfo {author}
  {\bibfnamefont{E.}~\bibnamefont{Gunn}}, \bibinfo {author}
  {\bibfnamefont{L.}~\bibnamefont{Yu}},\ and\ \bibinfo {author}
  {\bibfnamefont{B.}~\bibnamefont{Kahr}},\ }%
  \bibfield{journal}{%
  \Doi{10.1021/cg200640g}{\bibinfo {journal} {Crystal Growth \& Design}}\ }%
  \textbf{\bibinfo {volume} {11}},\ \bibinfo {pages} {4458} (\bibinfo {month}
  {Oct.}\ \bibinfo {year} {2011})\BibitemShut{NoStop}%
\bibitem{FujaraGeilSillescuFleischer1992}%
  \BibitemOpen
  \bibfield{author}{%
  \bibinfo {author} {\bibfnamefont{F.}~\bibnamefont{Fujara}}, \bibinfo {author}
  {\bibfnamefont{B.}~\bibnamefont{Geil}}, \bibinfo {author}
  {\bibfnamefont{H.}~\bibnamefont{Sillescu}},\ and\ \bibinfo {author}
  {\bibfnamefont{G.}~\bibnamefont{Fleischer}},\ }%
  \bibfield{journal}{%
  \Doi{10.1007/BF01323572}{\bibinfo {journal} {Zeitschrift Fur Physik
  B-condensed Matter}}\ }%
  \textbf{\bibinfo {volume} {88}},\ \bibinfo {pages} {195} (\bibinfo {month}
  {Aug.}\ \bibinfo {year} {1992})\BibitemShut{NoStop}%
\bibitem{CiceroneEdiger1996}%
  \BibitemOpen
  \bibfield{author}{%
  \bibinfo {author} {\bibfnamefont{M.~T.}\ \bibnamefont{Cicerone}}\ and\
  \bibinfo {author} {\bibfnamefont{M.~D.}\ \bibnamefont{Ediger}},\ }%
  \bibfield{journal}{%
  \Doi{10.1063/1.471433}{\bibinfo {journal} {Journal of Chemical Physics}}\ }%
  \textbf{\bibinfo {volume} {104}},\ \bibinfo {pages} {7210} (\bibinfo {month}
  {May}\ \bibinfo {year} {1996})\BibitemShut{NoStop}%
\bibitem{EdigerHarrowellYu2008}%
  \BibitemOpen
  \bibfield{author}{%
  \bibinfo {author} {\bibfnamefont{M.~D.}\ \bibnamefont{Ediger}}, \bibinfo
  {author} {\bibfnamefont{P.}~\bibnamefont{Harrowell}},\ and\ \bibinfo {author}
  {\bibfnamefont{L.}~\bibnamefont{Yu}},\ }%
  \bibfield{journal}{%
  \Doi{10.1063/1.2815325}{\bibinfo {journal} {Journal of Chemical Physics}}\ }%
  \textbf{\bibinfo {volume} {128}},\ \bibinfo {pages} {034709} (\bibinfo
  {month} {Jan.}\ \bibinfo {year} {2008})\BibitemShut{NoStop}%
\bibitem{StillingerDebenedetti2005}%
  \BibitemOpen
  \bibfield{author}{%
  \bibinfo {author} {\bibfnamefont{F.~H.}\ \bibnamefont{Stillinger}}\ and\
  \bibinfo {author} {\bibfnamefont{P.~G.}\ \bibnamefont{Debenedetti}},\ }%
  \bibfield{journal}{%
  \Doi{10.1021/jp0456584}{\bibinfo {journal} {Journal of Physical Chemistry
  B}}\ }%
  \textbf{\bibinfo {volume} {109}},\ \bibinfo {pages} {6604} (\bibinfo {month}
  {Apr.}\ \bibinfo {year} {2005})\BibitemShut{NoStop}%
\bibitem{KeithPadden1963}%
  \BibitemOpen
  \bibfield{author}{%
  \bibinfo {author} {\bibfnamefont{H.~D.}\ \bibnamefont{Keith}}\ and\ \bibinfo
  {author} {\bibfnamefont{F.~J.}\ \bibnamefont{Padden}},\ }%
  \bibfield{journal}{%
  \Doi{10.1063/1.1702757}{\bibinfo {journal} {Journal of Applied Physics}}\ }%
  \textbf{\bibinfo {volume} {34}},\ \bibinfo {pages} {2409} (\bibinfo {year}
  {1963})\BibitemShut{NoStop}%
\bibitem{BisaultRyschenkowFaivre1991}%
  \BibitemOpen
  \bibfield{author}{%
  \bibinfo {author} {\bibfnamefont{J.}~\bibnamefont{Bisault}}, \bibinfo
  {author} {\bibfnamefont{G.}~\bibnamefont{Ryschenkow}},\ and\ \bibinfo
  {author} {\bibfnamefont{G.}~\bibnamefont{Faivre}},\ }%
  \bibfield{journal}{%
  \Doi{10.1016/0022-0248(91)90647-N}{\bibinfo {journal} {Journal of Crystal
  Growth}}\ }%
  \textbf{\bibinfo {volume} {110}},\ \bibinfo {pages} {889} (\bibinfo {month}
  {Apr.}\ \bibinfo {year} {1991})\BibitemShut{NoStop}%
\bibitem{Tanaka2003}%
  \BibitemOpen
  \bibfield{author}{%
  \bibinfo {author} {\bibfnamefont{H.}~\bibnamefont{Tanaka}},\ }%
  \bibfield{journal}{%
  \Doi{10.1103/PhysRevE.68.011505}{\bibinfo {journal} {Physical Review E}}\ }%
  \textbf{\bibinfo {volume} {68}},\ \bibinfo {pages} {011505} (\bibinfo {month}
  {Jul.}\ \bibinfo {year} {2003})\BibitemShut{NoStop}%
\bibitem{KonishiTanaka2007}%
  \BibitemOpen
  \bibfield{author}{%
  \bibinfo {author} {\bibfnamefont{T.}~\bibnamefont{Konishi}}\ and\ \bibinfo
  {author} {\bibfnamefont{H.}~\bibnamefont{Tanaka}},\ }%
  \bibfield{journal}{%
  \Doi{10.1103/PhysRevB.76.220201}{\bibinfo {journal} {Physical Review B}}\ }%
  \textbf{\bibinfo {volume} {76}},\ \bibinfo {pages} {220201} (\bibinfo {month}
  {Dec.}\ \bibinfo {year} {2007})\BibitemShut{NoStop}%
\bibitem{SunXiChenEdigerYu2008}%
  \BibitemOpen
  \bibfield{author}{%
  \bibinfo {author} {\bibfnamefont{Y.}~\bibnamefont{Sun}}, \bibinfo {author}
  {\bibfnamefont{H.}~\bibnamefont{Xi}}, \bibinfo {author}
  {\bibfnamefont{S.}~\bibnamefont{Chen}}, \bibinfo {author}
  {\bibfnamefont{M.~D.}\ \bibnamefont{Ediger}},\ and\ \bibinfo {author}
  {\bibfnamefont{L.}~\bibnamefont{Yu}},\ }%
  \bibfield{journal}{%
  \Doi{10.1021/jp7120577}{\bibinfo {journal} {Journal of Physical Chemistry
  B}}\ }%
  \textbf{\bibinfo {volume} {112}},\ \bibinfo {pages} {5594} (\bibinfo {month}
  {May}\ \bibinfo {year} {2008})\BibitemShut{NoStop}%
\bibitem{CaroliCaroliRouletFaivre1989}%
  \BibitemOpen
  \bibfield{author}{%
  \bibinfo {author} {\bibfnamefont{B.}~\bibnamefont{Caroli}}, \bibinfo {author}
  {\bibfnamefont{C.}~\bibnamefont{Caroli}}, \bibinfo {author}
  {\bibfnamefont{B.}~\bibnamefont{Roulet}},\ and\ \bibinfo {author}
  {\bibfnamefont{G.}~\bibnamefont{Faivre}},\ }%
  \bibfield{journal}{%
  \Doi{10.1016/0022-0248(89)90623-4}{\bibinfo {journal} {Journal of Crystal
  Growth}}\ }%
  \textbf{\bibinfo {volume} {94}},\ \bibinfo {pages} {253} (\bibinfo {month}
  {Jan.}\ \bibinfo {year} {1989})\BibitemShut{NoStop}%
\bibitem{Eshelby1957}%
  \BibitemOpen
  \bibfield{author}{%
  \bibinfo {author} {\bibfnamefont{J.~D.}\ \bibnamefont{Eshelby}},\ }%
  \bibfield{journal}{%
  \bibinfo {journal} {Proc. Roy. Soc. London A}\ }%
  \textbf{\bibinfo {volume} {241}},\ \bibinfo {pages} {376 } (\bibinfo {year}
  {1957})\BibitemShut{NoStop}%
\bibitem{Slaughter2001}%
  \BibitemOpen
  \bibfield{author}{%
  \bibinfo {author} {\bibfnamefont{W.}~\bibnamefont{Slaughter}},\ }%
  \emph{\bibinfo {title} {{T}he {L}inearized {T}heory of {E}lasticity}},\
  \bibinfo {edition} {1st}\ ed.\ (\bibinfo {publisher} {Birkh\"auser},\
  \bibinfo {address} {Boston},\ \bibinfo {year} {2001})\BibitemShut{NoStop}%
\bibitem{MalandroLacks1999}%
  \BibitemOpen
  \bibfield{author}{%
  \bibinfo {author} {\bibfnamefont{D.~L.}\ \bibnamefont{Malandro}}\ and\
  \bibinfo {author} {\bibfnamefont{D.~J.}\ \bibnamefont{Lacks}},\ }%
  \bibfield{journal}{%
  \bibinfo {journal} {J. Chem. Phys.}\ }%
  \textbf{\bibinfo {volume} {110}},\ \bibinfo {pages} {4593} (\bibinfo {year}
  {1999})\BibitemShut{NoStop}%
\bibitem{MaloneyLemaitre2006}%
  \BibitemOpen
  \bibfield{author}{%
  \bibinfo {author} {\bibfnamefont{C.~E.}\ \bibnamefont{Maloney}}\ and\
  \bibinfo {author} {\bibfnamefont{A.}~\bibnamefont{Lema\^{\i}tre}},\ }%
  \bibfield{journal}{%
  \bibinfo {journal} {Phys. Rev. E}\ }%
  \textbf{\bibinfo {volume} {74}},\ \bibinfo {pages} {016118} (\bibinfo {year}
  {2006})\BibitemShut{NoStop}%
\bibitem{RodneySchuh2009a}%
  \BibitemOpen
  \bibfield{author}{%
  \bibinfo {author} {\bibfnamefont{D.}~\bibnamefont{Rodney}}\ and\ \bibinfo
  {author} {\bibfnamefont{C.}~\bibnamefont{Schuh}},\ }%
  \bibfield{journal}{%
  \bibinfo {journal} {Phys. Rev. Lett.}\ }%
  \textbf{\bibinfo {volume} {{102}}},\ \bibinfo {pages} {235503} (\bibinfo
  {year} {{2009}})\BibitemShut{NoStop}%
\bibitem{ChattorajCaroliLemaitre2011}%
  \BibitemOpen
  \bibfield{author}{%
  \bibinfo {author} {\bibfnamefont{J.}~\bibnamefont{Chattoraj}}, \bibinfo
  {author} {\bibfnamefont{C.}~\bibnamefont{Caroli}},\ and\ \bibinfo {author}
  {\bibfnamefont{A.}~\bibnamefont{Lemaitre}},\ }%
  \bibfield{journal}{%
  \Doi{10.1103/PhysRevE.84.011501}{\bibinfo {journal} {Physical Review E}}\ }%
  \textbf{\bibinfo {volume} {84}},\ \bibinfo {pages} {011501} (\bibinfo {month}
  {Jul.}\ \bibinfo {year} {2011})\BibitemShut{NoStop}%
\bibitem{SunXiEdigerYu2008}%
  \BibitemOpen
  \bibfield{author}{%
  \bibinfo {author} {\bibfnamefont{Y.}~\bibnamefont{Sun}}, \bibinfo {author}
  {\bibfnamefont{H.}~\bibnamefont{Xi}}, \bibinfo {author}
  {\bibfnamefont{M.~D.}\ \bibnamefont{Ediger}},\ and\ \bibinfo {author}
  {\bibfnamefont{L.}~\bibnamefont{Yu}},\ }%
  \bibfield{journal}{%
  \Doi{10.1021/jp709616c}{\bibinfo {journal} {Journal of Physical Chemistry
  B}}\ }%
  \textbf{\bibinfo {volume} {112}},\ \bibinfo {pages} {661} (\bibinfo {month}
  {Jan.}\ \bibinfo {year} {2008})\BibitemShut{NoStop}%
\bibitem{NaokiKoeda1989}%
  \BibitemOpen
  \bibfield{author}{%
  \bibinfo {author} {\bibfnamefont{M.}~\bibnamefont{Naoki}}\ and\ \bibinfo
  {author} {\bibfnamefont{S.}~\bibnamefont{Koeda}},\ }%
  \bibfield{journal}{%
  \Doi{10.1021/j100339a078}{\bibinfo {journal} {Journal of Physical
  Chemistry}}\ }%
  \textbf{\bibinfo {volume} {93}},\ \bibinfo {pages} {948} (\bibinfo {month}
  {Jan.}\ \bibinfo {year} {1989})\BibitemShut{NoStop}%
\bibitem{BuchenauWischnewski2004}%
  \BibitemOpen
  \bibfield{author}{%
  \bibinfo {author} {\bibfnamefont{U.}~\bibnamefont{Buchenau}}\ and\ \bibinfo
  {author} {\bibfnamefont{A.}~\bibnamefont{Wischnewski}},\ }%
  \bibfield{journal}{%
  \Doi{10.1103/PhysRevB.70.092201}{\bibinfo {journal} {Physical Review B}}\ }%
  \textbf{\bibinfo {volume} {70}},\ \bibinfo {pages} {092201} (\bibinfo {month}
  {Sep.}\ \bibinfo {year} {2004})\BibitemShut{NoStop}%
\bibitem{NovikovSokolov2004}%
  \BibitemOpen
  \bibfield{author}{%
  \bibinfo {author} {\bibfnamefont{V.~N.}\ \bibnamefont{Novikov}}\ and\
  \bibinfo {author} {\bibfnamefont{A.~P.}\ \bibnamefont{Sokolov}},\ }%
  \bibfield{journal}{%
  \Doi{10.1038/nature02947}{\bibinfo {journal} {Nature}}\ }%
  \textbf{\bibinfo {volume} {431}},\ \bibinfo {pages} {961} (\bibinfo {month}
  {Oct.}\ \bibinfo {year} {2004})\BibitemShut{NoStop}%
\bibitem{PlazekBeroChay1994}%
  \BibitemOpen
  \bibfield{author}{%
  \bibinfo {author} {\bibfnamefont{D.~J.}\ \bibnamefont{Plazek}}, \bibinfo
  {author} {\bibfnamefont{C.~A.}\ \bibnamefont{Bero}},\ and\ \bibinfo {author}
  {\bibfnamefont{I.~C.}\ \bibnamefont{Chay}},\ }%
  \bibfield{journal}{%
  \Doi{10.1016/0022-3093(94)90431-6}{\bibinfo {journal} {Journal of
  Non-crystalline Solids}}\ }%
  \textbf{\bibinfo {volume} {172}},\ \bibinfo {pages} {181} (\bibinfo {month}
  {Sep.}\ \bibinfo {year} {1994})\BibitemShut{NoStop}%
\bibitem{SunXiEdigerRichertYu2009}%
  \BibitemOpen
  \bibfield{author}{%
  \bibinfo {author} {\bibfnamefont{Y.}~\bibnamefont{Sun}}, \bibinfo {author}
  {\bibfnamefont{H.}~\bibnamefont{Xi}}, \bibinfo {author}
  {\bibfnamefont{M.~D.}\ \bibnamefont{Ediger}}, \bibinfo {author}
  {\bibfnamefont{R.}~\bibnamefont{Richert}},\ and\ \bibinfo {author}
  {\bibfnamefont{L.}~\bibnamefont{Yu}},\ }%
  \bibfield{journal}{%
  \Doi{10.1063/1.3200228}{\bibinfo {journal} {Journal of Chemical Physics}}\ }%
  \textbf{\bibinfo {volume} {131}},\ \bibinfo {pages} {074506} (\bibinfo
  {month} {Aug.}\ \bibinfo {year} {2009})\BibitemShut{NoStop}%
\bibitem{AngellNgaiMcKennaMcMillanMartin2000}%
  \BibitemOpen
  \bibfield{author}{%
  \bibinfo {author} {\bibfnamefont{C.~A.}\ \bibnamefont{Angell}}, \bibinfo
  {author} {\bibfnamefont{K.~L.}\ \bibnamefont{Ngai}}, \bibinfo {author}
  {\bibfnamefont{G.~B.}\ \bibnamefont{McKenna}}, \bibinfo {author}
  {\bibfnamefont{P.~F.}\ \bibnamefont{McMillan}},\ and\ \bibinfo {author}
  {\bibfnamefont{S.~W.}\ \bibnamefont{Martin}},\ }%
  \bibfield{journal}{%
  \bibinfo {journal} {J. Appl. Phys.}\ }%
  \textbf{\bibinfo {volume} {88}},\ \bibinfo {pages} {3113} (\bibinfo {year}
  {2000})\BibitemShut{NoStop}%
\bibitem{DebenedettiStillinger2001}%
  \BibitemOpen
  \bibfield{author}{%
  \bibinfo {author} {\bibfnamefont{P.~G.}\ \bibnamefont{Debenedetti}}\ and\
  \bibinfo {author} {\bibfnamefont{F.~H.}\ \bibnamefont{Stillinger}},\ }%
  \bibfield{journal}{%
  \Doi{10.1038/35065704}{\bibinfo {journal} {Nature}}\ }%
  \textbf{\bibinfo {volume} {410}},\ \bibinfo {pages} {259} (\bibinfo {month}
  {Mar.}\ \bibinfo {year} {2001})\BibitemShut{NoStop}%
\bibitem{Note1}%
  \BibitemOpen
  \bibinfo {note} {The fact that interfacial stress does not increase
  indefinitely as growth proceeds is consistent with the absence of free-volume
  creation.}\BibitemShut{Stop}%
\bibitem{Note2}%
  \BibitemOpen
  \bibinfo {note} {The characteristic time scale for the transient dynamics is
  of course the matrix relaxation time $\tau _\alpha $}\BibitemShut{NoStop}%
\end{thebibliography}
%

\end{document}